 \documentclass[twocolumn,showpacs,preprintnumbers,amsmath,amssymb, superscriptaddress]{revtex4}
 \usepackage{graphicx}
 \usepackage{dcolumn}
 \usepackage{bm}
 \usepackage{color,ulem}

\begin{document}

\title{Near-Field Resonance at Far-Field Anti-Resonance: Plasmonically Enhanced Light Emission with Minimum Scattering Nanoantennas  }

  \author{S. R. K. Rodriguez}\email{s.rodriguez@amolf.nl}
  \address{Center for Nanophotonics, FOM Institute AMOLF, c/o Philips Research Laboratories, High Tech Campus 4, 5656 AE Eindhoven, The Netherlands}

   \author{O. T. A. Janssen}
   \address{Optics Research Group, Delft University of Technology, 2628 CJ Delft, The Netherlands}

  \author{G. Lozano}
  \address{Center for Nanophotonics, FOM Institute AMOLF, c/o Philips Research Laboratories, High Tech Campus 4, 5656 AE Eindhoven, The Netherlands}

  \author{A. Omari}
  \address{Center for Nano and Biophotonics, Ghent University, Belgium.}
  \address{Physics and Chemistry of Nanostructures, Ghent University,  Belgium}

  \author{Z. Hens}
  \address{Center for Nano and Biophotonics, Ghent University, Belgium.}
  \address{Physics and Chemistry of Nanostructures, Ghent University,  Belgium}

  \author{J. G\'{o}mez Rivas}
  \address{Center for Nanophotonics, FOM Institute AMOLF, c/o Philips Research Laboratories, High Tech Campus 4, 5656 AE Eindhoven, The Netherlands}
  \address{COBRA Research Institute, Eindhoven University of Technology, P.O. Box 513, 5600 MB Eindhoven, The Netherlands}
    \date{\today}

\begin{abstract}
We demonstrate that a periodic array of optical antennas sustains a
resonant Near-Field (NF) and an anti-resonant Far-Field (FF) at the
same energy and in-plane momentum. This phenomenon arises in the
context of coupled plasmonic lattice resonances, whose bright and
dark character is interchanged at a critical antenna length. The
energies of these modes anti-cross in the FF, but cross in the NF.
Hence, we observe an extremely narrow bandwidth emission enhancement
from quantum dots in the proximity of the array, while the antennas
scatter minimally into the FF. Simulations reveal that a standing
wave with a quadrupolar field distribution is the origin of this
dark collective resonance.

 \end{abstract}

\maketitle

 Resonance phenomena are at the heart of nearly any approach towards
controlling the emission, manipulation, or detection of light. A
resonance frequency is commonly identified as that for which the
response of a system is enhanced. For a plane wave excitation, this
implies that more energy is removed from an incident beam at
resonance than off resonance. In the presence of surface
electromagnetic waves, the definition of a resonance requires a
careful reconsideration, since the spectrum of radiation may be
different in the Near-Field (NF) with respect to the Far-Field
(FF)~\cite{Greffet00, Bryant08, Lee09}. Such differences have a
profound significance for the field of metallic nano-optics, partly
because the emergence of optical antennas has been largely fueled by
an increased capability to manipulate optical near
fields~\cite{Novotny11}. The basis for achieving such a control
relies on the excitation of surface plasmon polaritons - collective
oscillations of the free electrons in the metal driven by an
electromagnetic field. However, not all surface plasmon modes are
visible in the FF spectrum. In 2001, a class of plasmonic  modes
which manifest as resonances that can be excited or observed in the
NF but not in the FF, so-called dark modes,  were theoretically
predicted~\cite{Stockman01}. These dark resonances have
attracted much interest in recent years~\cite{Alivisatos,
Stockman10}, as they hold supreme qualities for the realization of
SPASERS~\cite{Stockman03, Stockman05}, subwavelength guiding of
optical radiation with suppressed radiative losses ~\cite{Liu09},
plasmonic analogs of electromagnetically induced
transparency~\cite{Zhang08, Yannopapas09, Giessen09,
Kekatpure&Brongersma10}, and cloaked sensors~\cite{Alu09}. In
particular, theoretical work by Al\`u and Engheta suggests that it
is possible to strongly suppress the FF radiation from a nanoantenna
while preserving an enhanced NF sensitivity~\cite{Alu09, Alu10} - an
optical counterpart to radio frequency minimum scattering antennas.
As pointed out by Garc\'\i a de Abajo in Ref.~\cite{Garcia09}, by
having an enhanced interaction with its local environment but a
minimum interaction with distant sources and detectors, such a
minimum scattering antenna is effectively seeing without being seen.

In this Letter, we experimentally demonstrate that a periodic array
of plasmonic nanoantennas displays a local minimum in its FF
extinction but a maximum in the average NF enhancement in the plane
of the array. This is evidenced from an enhanced emission of quantum
dots in the vicinity of the array at an energy and in-plane momentum
for which the FF extinction is minimized. Hence, a NF resonance
co-exists in energy and in-plane momentum with a FF anti-resonance. Finite
Difference in Time Domain (FDTD) simulations show that this
phenomenon is based on the coupling of two counter-propagating
surface polaritons whose energies anti-cross in the FF extinction but cross
in the NF enhancement. A standing wave with a quadrupolar, i.e.,
dipole forbidden, field distribution is formed at the NF crossing
energy. This results in a collective resonance which is dark in the
FF, but bright in the NF.

\begin{figure}
\centerline{\scalebox{0.47}{\includegraphics{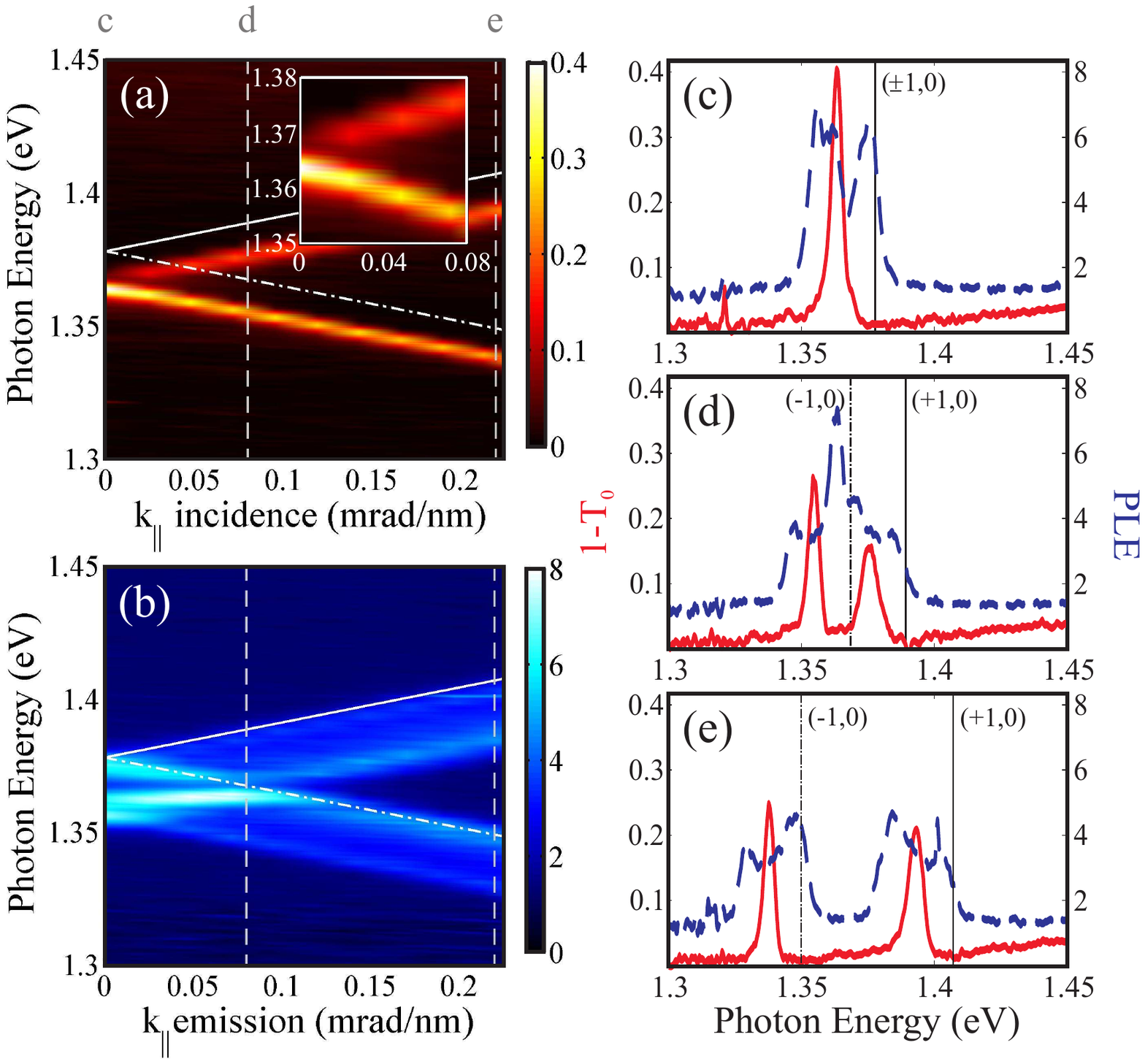}}}
\caption{Experimental dispersion relations in (a) extinction, and
(b) PhotoLuminescence Enhancement (PLE), for the nanoantenna array
described in the text. The photon energy and wave vector component
parallel to surface refer to the (a) incident and (b) emitted
radiation. The solid and dash-dotted white lines indicate the (+1,0)
and (-1,0) Rayleigh anomalies, respectively. The inset in (a) shows
a magnified view of the low $k_{||}$ region, where the gap opens.
The extinction is shown as a solid red line and the PLE as a dashed blue
line for (c) $k_{||}=0$, (d) $k_{||}= 0.08$ mrad/nm, and (e) $k_{||}= 0.22$ mrad/nm; the latter two cases are indicated
by the dashed lines in (a) and (b).} \label{fig:fig1}
\end{figure}

For the experiments, we have fabricated a $2\times 2$ mm$^2$
periodic array of gold nanoantennas  onto a silica substrate by
means of electron beam lithography. The antennas have dimensions
$270\times 80\times 40$~nm$^3$, and the lattice constants are
$a_x=600$ nm and $a_y=300$ nm. We spin-coated a 600 nm layer of
PbS/CdS core/shell Quantum Dots (QDs) in a polystyrene matrix,
hereafter referred as the QD layer, on top of the array. The QDs
emit at a peak energy of $1.33$ eV with a full width at half maximum
of $0.28$ eV. Further details on their synthesis and optical
properties may be found in the supplemental
information~\cite{supp_QD}.

Figure \ref{fig:fig1}(a) shows the measured variable angle FF
extinction, defined as $1-T_0$, with $T_0$ the zeroth order
transmission through the sample normalized to the transmission
through the substrate and QD layer. The measurements are shown as  a
function of the incident photon energy and the  wave vector
component parallel to the long axis of the antennas, $k_{||}= k_0
\sin(\theta_{in}) \hat{x}$, where $k_0 = \frac{2 \pi }{\lambda_0}$
is the magnitude of the free space wave vector and $\theta_{in}$ is
the angle of incidence. The incident light is polarized parallel to
the short axis of the nanoantennas, i.e., s-polarized. The ($+1,0$)
and ($-1,0$) Rayleigh anomalies are indicated  by the white solid
and dash-dotted lines, respectively. They represent the conditions for
which the corresponding diffraction orders are radiating in the
plane of the array. Their dispersion is calculated from the
conservation of the parallel component of the wave vector, expressed
as $k_{out}^2 = ( k_{x} \pm  m_1 G_x)^2 + ( k_{y} \pm  m_2 G_y)^2$,
with $k_{out}$ the magnitude of the scattered wave vector, $k_\| =
(k_{x}, k_{y})$ the wave vector components parallel to the surface,
the integers ($m_1$, $m_2$) defining the order of diffraction, and
$\vec{G} = (G_x= \frac{2 \pi}{a_x}$,$G_y = \frac{2 \pi}{a_y}$) the
reciprocal lattice vector. An effective refractive index of n=1.50
due to the underlying substrate and the QD layer was used to
calculate the Rayleigh anomalies. The two peaks in extinction
following the dispersion of the Rayleigh anomalies on the low energy
side correspond to the excitation of Surface Lattice Resonances
(SLRs), which are collective Fano resonances arising from the
diffractive coupling of localized surface
plasmons~\cite{Zou&Schatz04b, Kravets08, Auguie&Barnes08, Crozier08,
Vecchi09, Vecchi09b, Giannini10}. The mutual coupling of SLRs leads
to an anti-crossing in their dispersion relation, i.e., the opening
of a frequency gap close to $k_{||}= 0$~\cite{SRKR11a}. The inset of
Fig.~\ref{fig:fig1}(a) displays a magnified view of the gap. Notice
that the dispersion of the bright (-1,0) SLR flattens and the
extinction increases near the anti-crossing. This indicates the
formation of standing waves with an enhanced density of optical
states at the band edge. A narrowing linewidth and diminishing
extinction are observed for the (+1,0) SLR, which are the signature
of subradiant damping~\cite{Ropers, SRKR11a}.

Figure~\ref{fig:fig1}(b) shows the measured variable angle
PhotoLuminescence Enhancement (PLE), defined as $I_{in}/I_{out}$, with
$I_{in}$ and $I_{out}$  the emission from the QD layer inside and
outside the nanoantenna array, respectively. The sample was excited
by the 514 nm line of an Ar/Kr laser at a fixed angle ($9^{\circ}$)
of incidence, and the s-polarized emission was collected as a
function of the angle $\theta_{em}$ that the detector subtends with
respect to the normal. The PLE is shown as a function of the emitted
photon energy and wave vector component parallel to the long axis of
the antennas, $k_\|= k_0 \sin(\theta_{em}) \hat{x}$. Although the
PLE displays features approximately resembling those in extinction
(albeit broader), a detailed inspection reveals an unusual
non-correspondence for some values of $k_{||}$. To illustrate this,
we plot in Figs.~\ref{fig:fig1}(c)-(e) the spectra in Figs.~\ref{fig:fig1}(a)
and~\ref{fig:fig1}(b) at three values of $k_{||}$, with the extinction as solid red lines and
the PLE as dashed blue lines, respectively. At $k_{||}=0$
(Fig.~\ref{fig:fig1}(c)), a single peak in extinction associated
with the bright (-1,0) SLR arises at $1.363$ eV. The PLE displays
two peaks with a $\sim 6$-fold enhancement: one at $1.36 $ eV
corresponding to the bright (-1,0) SLR, and another at $1.375$ eV
corresponding to the dark (+1,0) SLR. We attribute the enhancement
of the QD emission by the (+1,0) SLR, which is dark in the FF, to
the local nature of the excitation. As it was shown in a recent
work~\cite{SRKR11a}, the dark and bright character of SLRs has its
origin in the symmetry of the modes. However, the QD emitters are
sensitive to the local field rather than the global symmetry of the
array, which allows for their interaction with dark
modes~\cite{Stockman01}. Figure~\ref{fig:fig1}(d) shows the spectra
at $k_{||}=0.08$ mrad/nm, where the extinction displays peaks at
$1.376$ eV and $1.354$ eV corresponding to the (+1,0) and (-1,0)
SLRs, respectively. The PLE also shows two broad features centered
at approximately the same energies, but more striking is the narrow
(60 meV full width at half maximum) feature at $1.364$ eV leading to
a $\sim7$-fold enhancement of the QD emission. Notice that at the
same energy the FF extinction is very low. It is remarkable that at
an energy and in-plane momentum for which the nanoantennas scatter
minimally when probed from the FF  there is an emission enhancement
that is superior to the one observed for the bright mode at
$k_{||}=0$. The physics behind this NF resonance at a FF
anti-resonance is the focus of this work. For comparison, we show in
Fig.~\ref{fig:fig1}(e) the measurements at $k_{||}=0.22$ mrad/nm,
where two broad features in the PLE are observed near the energies
of the ($\pm 1,0$) SLRs in extinction, as previous work has
shown~\cite{Vecchi09}.

\begin{figure}
\centerline{\scalebox{0.48}{\includegraphics{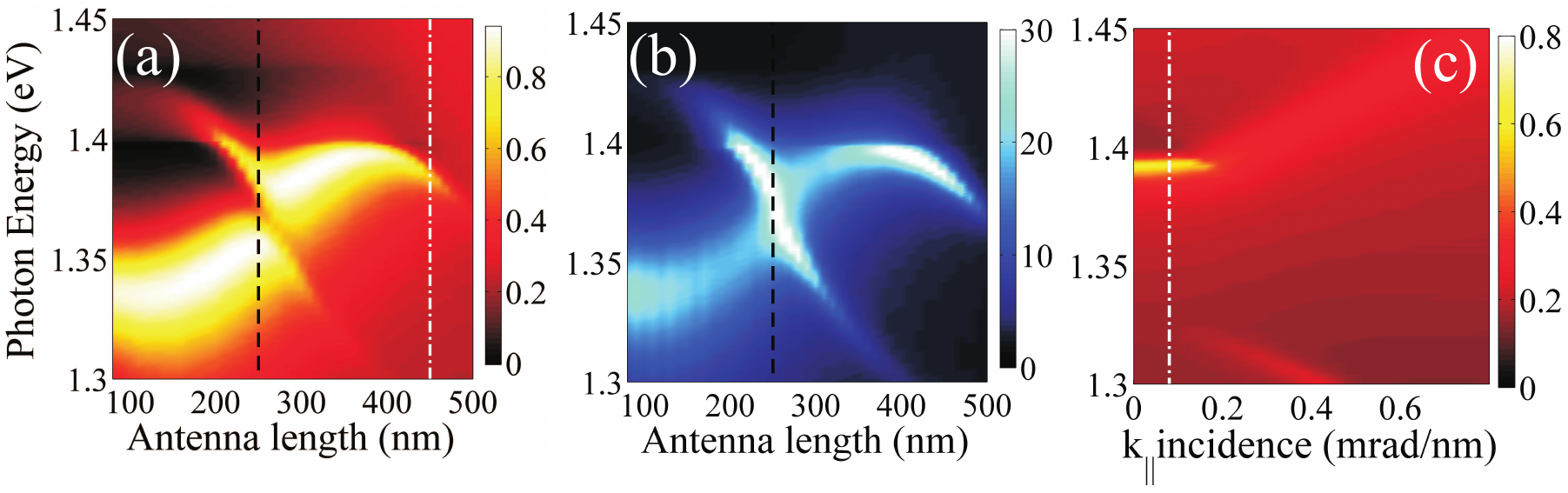}}}
\caption{ FDTD spectra in (a) extinction, and (b) NF intensity
enhancement at a plane intersecting the antennas at their
mid-height, for arrays of antennas with width =110 nm, height = 40
nm, and variable lengths. The arrays are illuminated by a plane wave
with $k_\parallel= 0.08$ mrad/nm. (c) Dispersion relation in
extinction for an array of antennas with dimensions $450 \times 110
\times 40$ nm$^3$. The black dashed lines in (a) and (b) indicate
the value of the antenna length investigated in Fig.~\ref{fig:fig2}(c).
The white dash-dotted lines in both (a) and (c) indicate
the extinction for an array of antennas of length = 450 nm, illuminated by a plane wave with $k_\parallel= 0.08$
mrad/nm.} \label{fig:fig2}
\end{figure}

In what follows, we elucidate by means of FDTD simulations the
conditions leading to the NF resonance at a FF anti-resonance. We
investigate gold nanoantenna arrays with the same lattice constants
as in the experiments, surrounded by a fully homogeneous environment
of n=1.46. The dielectric function of gold is taken from
Ref.~\cite{Palik91}, and fitted in the range of interest with a
Drude model. The incident light has an in-plane momentum and a
polarization vector as in the experiments.

Figure~\ref{fig:fig2} shows simulations results for the (a) FF
extinction and (b) average NF Intensity Enhancement (IE) defined as
IE $= |E|^2/|E_0|^2$, with $E$ the total electric field at a plane
intersecting the antennas at their mid-height and $E_0$ the incident
field. The antenna dimensions are $L \times 110\times 40$~nm$^3$,
and the spectral response to a plane wave with  $k_{||} =
0.08$~mrad/nm is shown as a function of the antenna length $L$.
Besides $L$, all other parameters are kept constant. The high and
low energy resonances in both figures are the ($+1,0$) and ($-1,0$)
SLRs, respectively. Their extinction and NF IE
vary with the antenna length as a consequence of retardation and
radiative damping. Notice that for $L \gtrsim 250$ nm the (+1,0) SLR
is bright, i.e., it has a high extinction, whereas the (-1,0) SLR is
dark, i.e., it has a narrowing linewidth and a low extinction. The
extinction of the (-1,0) SLR vanishes for $L \gtrsim 350$ nm. To
exemplify the FF dispersion diagram in the long antenna regime, we
show in Fig.~\ref{fig:fig2}(c) the variable angle extinction spectra
calculated for an array of antennas with dimensions $450 \times
110\times 40$~nm$^3$. We clarify that the spectra in
Fig.~\ref{fig:fig2}(c) at $k_{||} = 0.08$ mrad/nm and in
Fig.~\ref{fig:fig2}(a) at $L = 450$ nm (both denoted by a white
dash-dotted line) are identical. The dispersion diagram in
Fig.~\ref{fig:fig2}(c) shows that, as the two SLR bands approach
each other near normal incidence, the high energy band flattens
and the extinction and linewidth of the low energy band vanish.
This results in the opening of a large gap. As observed in
Fig.~\ref{fig:fig2}(a) and ~\ref{fig:fig2}(b), the properties of the
high and low energy SLRs bands are interchanged  for $L \lesssim
250$ nm. Thus, in the short antenna regime the flattening of the
band occurs for the (-1,0) SLR, while subradiant damping onsets for
the (+1,0) SLR. This explains why in a recent work the (+1,0) SLR
was bright and the (-1,0) SLR was dark ~\cite{SRKR11a} (as in
Fig.~\ref{fig:fig2}(c)), whereas in the measurements presented here
(Fig.~\ref{fig:fig1}) this behavior is reversed. These results
demonstrate that the onset of subradiant damping can be modified, or
even halted for a given band, by designing the antenna length.
Furthermore, a remarkable phenomenon arises at the critical antenna
length for which the properties of SLRs are swapped, which  occurs
for $230$ nm $\lesssim  L \lesssim 270$ nm.
In this regime, the FF displays an energy anti-crossing
characteristic of coupled surface modes, but the NF displays a
crossing of the two bands. At this point the NF is resonant at an
energy and in-plane momentum for which the FF is anti-resonant.

\begin{figure}
\centerline{\scalebox{0.46}{\includegraphics{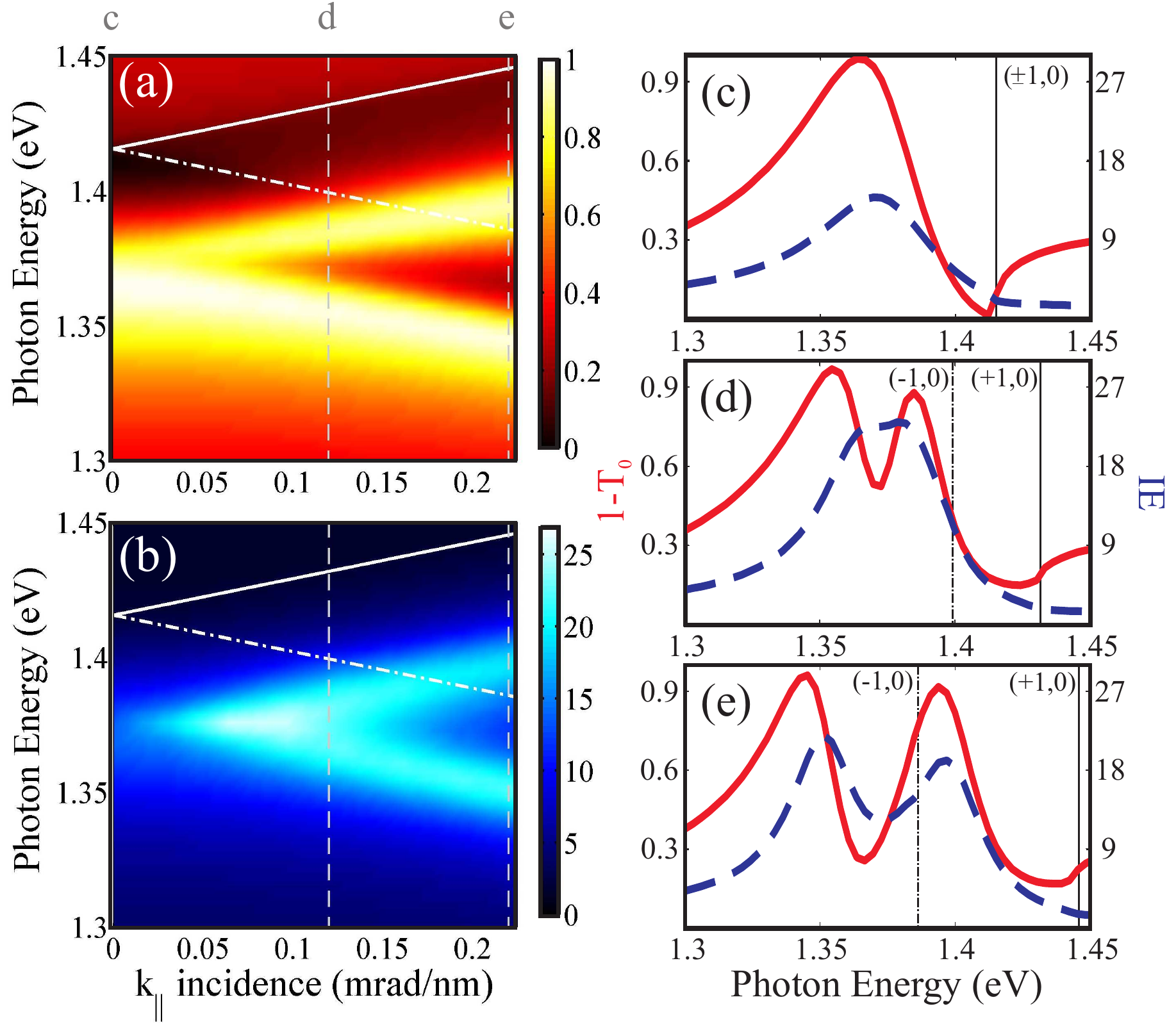}}}
\caption{(a) Extinction and (b) average NF Intensity Enhancement
(IE) at a plane at the mid-height of the antennas, which have
dimensions $250 \times 110 \times 40$ nm$^3$ . The solid and
dash-dotted white lines indicate the (+1,0) and (-1,0) Rayleigh
anomalies, respectively. The extinction is shown as a solid red line and the NF IE as a dashed blue
line for (c) $k_{||}=0$, (d) $k_{||}= 0.12$ mrad/nm, and (e) $k_{||}= 0.22$ mrad/nm; the latter two cases are indicated
by the dashed lines in (a) and (b).} \label{fig:fig3}
\end{figure}

The connection between the above spectra as a function of the
antenna length $L$ and the experimental results of
Fig.~\ref{fig:fig1} is elucidated in Fig.
\ref{fig:fig3} by considering FDTD simulations for antennas of $L=250$ nm. This length
is indicated by the black dashed lines in Figs.~\ref{fig:fig2}(a) and
~\ref{fig:fig2}(b), and it is very close to the one in the
experiments. The value of $L$ was adjusted slightly to obtain resonances at the same energies as
 in the experiment, since differences in geometry (especially near the antenna corners) and in the dielectric function of gold result in deviations from experiments. All other parameters in the simulation are identical to
the ones discussed for Fig.~\ref{fig:fig2}, except that now the
spectra are calculated as function of the incident $k_{||}$ for a
fixed antenna length. The FF extinction is shown in
Fig.~\ref{fig:fig3}(a), and the NF IE is shown in
Fig.~\ref{fig:fig3}(b). As in the experiments,  the ($\pm1$,0) SLRs
are red-shifted with respect to the Rayleigh anomalies in
extinction. The NF IE displays similar features, but they are broader, and shifted towards higher $k_{||}$ with respect to the FF. The latter condition leads to a maximum NF IE at a nonzero $k_{||}$, as we show next by
making cuts of Figs.~\ref{fig:fig3}(a) and~\ref{fig:fig3}(b) at three
values of $k_{||}$. At normal
incidence, Fig.~\ref{fig:fig3}(c) shows a single extinction peak
attributed to ($\pm 1, 0$) SLRs. The NF also displays a single peak
because the dark mode can not be excited by the plane wave due to
its antisymmetric character~\cite{SRKR11a}. In
Fig.~\ref{fig:fig3}(d) we consider $k_{||}=0.12$ mrad/nm, where the
NF displays a single peak at an energy for which the FF displays a
dip between two peaks. At this energy a NF resonance coincides with
a FF anti-resonance. Note that the same behavior can be observed for
$k_{||}=0.08$ mrad/nm (the value inspected in the experiments), but
with a more intense NF at the expense of a shallower extinction dip.
This trade-off between NF enhancement and diminished extinction has
its origin in absorption by the metal.  As we will show ahead,
destructive interference may minimize the FF scattering, but the
intensified NF increases the absorption and therefore sets a lower
limit on the extinction. In view of this, it is not entirely clear
to us why the extinction is significantly lower in the experiments
than in the simulations. However, a qualitative agreement is
observed, whereby the PLE is intensified where the response to the
plane wave is minimized. Finally, Fig.~\ref{fig:fig3}(e) shows the
spectra at $k_{||}=0.22$ mrad/nm, where two distinct resonances are
observed in both the FF and the NF. By comparing Figs. \ref{fig:fig3}(d)
and~\ref{fig:fig3}(e) it can be recognized that the single NF peak
in Fig. 3(d) corresponds to a point of degeneracy. The (+1,0) SLR is
red-shifted and the (-1,0) SLR is blue-shifted with respect to the
FF  such that their energies are equal in the NF. The origin of this
degeneracy can be traced to the degeneracy of the ($\pm 1,0$)
Rayleigh anomalies at normal incidence, which is linearly lifted as
$k_{||}$ increases.

\begin{figure}
\centerline{\scalebox{0.48}{\includegraphics{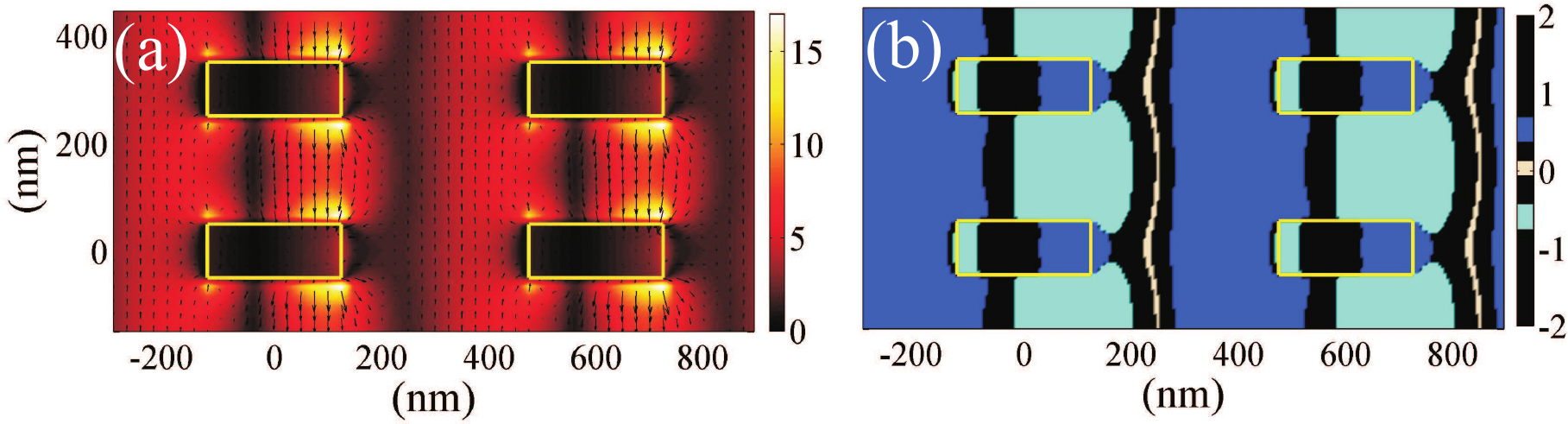}}}
\caption{(a) Field enhancement (in color scale) and real part of the
scattered field at an arbitrary phase (arrows), and (b) phase
difference between the scattered and incident field in units of
$\pi$. Both (a) and (b) are calculated at a plane intersecting the
antennas (delimited by the yellow lines) at their mid-height. The
antenna array is identical to the one described for Fig. 3. The
illuminating plane wave has an energy of 1.37 eV and $k_{||}=
0.12$~mrad/nm, corresponding to the NF resonance at a FF anti-resonance in Fig. 3(d).} \label{fig:fig4}
\end{figure}

The suppressed FF scattering and simultaneous intense NF  can be
understood by considering the electric field amplitude and phase
response of the array. This is shown in Fig.~\ref{fig:fig4} for the
energy of the dip in extinction in Fig.~\ref{fig:fig3}(d).
Figure~\ref{fig:fig4}(a) shows in color scale the total electric
field enhancement, i.e., $|E|/|E_0|$. The scattered field is
represented by the arrows. The 4 hot-spots
near the corners of the antennas and the resultant field pattern indicate
that the excited mode has a quadrupolar character, i.e., it is
dipole forbidden. However, the enhancement on the right side of each
antenna is larger than on the left side. This results in a
non-vanishing dipole moment, allowing the excitation of this mode
and the finite extinction observed in Fig.~\ref{fig:fig3}(d).
Figure~\ref{fig:fig4}(b) shows the differences in phase between the
scattered and incident fields, i.e., $\phi_{sca} - \phi_{inc}$. Two
values, which are ($-0.6 \pm 0.15$) $\pi$ (light blue) and ($0.5 \pm
0.15$) $\pi$ (dark blue), are predominant throughout space. Their
difference, $1.1 \pi$, is very close the out of phase condition.
Therefore, we observe in Fig.~\ref{fig:fig4}(b) that $\phi_{sca} -
\phi_{inc}$ inside the antennas has a difference of $\sim \pi$ with
respect to $\phi_{sca} - \phi_{inc}$ in the dielectric surroundings,
resulting in a  collective suppression of the FF scattering by
destructive interference. Furthermore, this spatial distribution of
the phase corresponds to the formation of a standing wave, arising
from the interaction of two counter-propagating surface polaritons in the
plasmonic crystal. This can be observed in the movies of the total
electric field provided in the supplemental
information~\cite{supp_movies}, where two counter-propagating surface polaritons are observed at the high and low
energy extinction peaks, and  a quadrupolar standing wave is observed at the extinction
dip, all referring to the spectrum in Fig.~\ref{fig:fig3}(d). These results elucidate
the nature of a minimum scattering antenna array with an enhanced
local field, which can be seen by the QDs emission without being
seen in the FF.

In conclusion, we have demonstrated that differences in the
far-field and near-field spectra of radiation of optical antenna
arrays can be exploited to create a medium with large local field
enhancements but minimized extinction. This local field enhancement
allowed us to observe an extremely narrow bandwidth emission
enhancement of quantum dots in the vicinity of the array with a
simultaneous minimum in extinction. Dark plasmonic resonances hold
remarkable features for modified light emission and sensing, both of
which depend on the local field rather than on the global symmetry
of the array determining the FF extinction. Our results demonstrate
that dipole inactive or dark modes are capable of enhancing the
performance of photonic devices to a level comparable, or even
superior, to bright modes.

This work was supported by the Netherlands Foundation Fundamenteel
Onderzoek der Materie (FOM) and the Nederlandse Organisatie voor
Wetenschappelijk Onderzoek (NWO), and is part of an industrial
partnership program between Philips and FOM. O. T. A. J.
acknowledges support from the Dutch Technology Foundation STW, which
is the applied science division of NWO, and the Technology Programme
of the Ministry of Economic Affairs (project number 10301). A. O.
acknowledges the Institute for the Promotion of Innovation through
Science and Technology in Flanders (IWT-Vlaanderen). Z. H.
acknowledges BelSPo (IAP 6.10, photonics@be) and the FWO-Vlaanderen
(project nr. G.0794.10) for research funding.


\end{document}